\documentclass[aps]{revtex4}
\usepackage{amsfonts}
\usepackage{amsmath}
\usepackage{amssymb,epsf}
\usepackage{color}
\usepackage{graphicx}
\usepackage{epstopdf}
\usepackage{float}
\usepackage{caption}
\usepackage{subfig}

\begin{document}

\title{Stable three-dimensional (un)charged AdS gravastars in gravity's
rainbow}
\author{H. Barzegar$^{1}$, M. Bigdeli$^{1}$ \footnote{%
email address: m\_bigdeli@znu.ac.ir}, G. H. Bordbar$^{2,3}$ \footnote{%
email address: ghbordbar@shirazu.ac.ir}, and B. Eslam Panah$^{4,5,6}$
\footnote{%
email address: eslampanah@umz.ac.ir}}

\begin{abstract}
In this work, we study the three-dimensional AdS gravitational vacuum stars
(gravastars) in the context of gravity's rainbow theory. Then we extend it
by adding the Maxwell electromagnetic field. We compute the physical
features of gravastars, such as proper length, energy, entropy, and junction
conditions. Our results show that the physical parameters for charged and
uncharged states depend significantly on rainbow functions. Besides from
charged state, they also depend on the electric field. Finally, we explore
the stability of thin shell of three-dimensional (un)charged AdS gravastars
in gravity's rainbow. We show that the structure of thin shell of these
gravastars may be stable and is independent of the type of matter.
\end{abstract}

\pacs{ 04.70.-s, 04.30.-w}
\keywords{garavastar, BTZ black hole, gravity's rainbow}

\address{ $^{1}$ Physics Department, College of Sciences, University of Zanjan, Zanjan, Iran\\	
$^{2}$ Physics Department and Biruni Observatory, Shiraz University, Shiraz 71454, Iran\\
$^{3}$ Department of Physics and Astronomy, University of Waterloo, 200 University Avenue West, Waterloo, Ontario N2L3G1, Canada\\
$^{4}$  Department of Theoretical Physics, Faculty of Science, University of Mazandaran, P. O. Box 47415-416, Babolsar, Iran\\
$^{5}$ ICRANet-Mazandaran, University of Mazandaran, P. O. Box 47415-416, Babolsar, Iran\\
$^{6}$ ICRANet, Piazza della Repubblica 10, I-65122 Pescara, Italy}

\maketitle

\section{Introduction}

Mazur and Motola first proposed the notion of a gravitational vacuum star
(gravastar) as an alternative to the black hole \cite{1,2}. In other words,
gravastars are free of any kind of singularity. The gravastar consists of
three regions as follows \cite%
{Visser2004,Bilic2006,Cattoen2005,Carter2005,DeBenedictis2006,Lobo2007,Horvat2007,Chirenti2007,Rocha2008,Horvat2009}%
. The interior region ($0<r<r_{1}$) has a vacuum explain by the dark energy,
with equation of state (EOS) $p=-\rho $. Shell region ($r_{1}<r<r_{2} $ )
consists of ultra-stiff perfect fluid with EoS $p=\rho $ that first
reported, by Zeldovich and Novikov in context of cosmology \cite{3,3a}. The
exterior region ($r>r_{2}$ ) is a vacuum region with EOS $p=0$ which can be
described by Schwarzschild, Kerr and Kerr-Neumann, Reissner-Nordstr\"{o}m
metrics as requested under the situation. Commonly described the
Schwarzschild \cite{4,5} and Reissner-Nordstr\"{o}m metric \cite{6,7}
depending the gravastar is charged and rotating or not.

In 1992, B\~{a}nados, Teitelboim, and Zanelli proposed a new black hole
solution by considering three-dimensional gravity with negative cosmological
constant \cite{8}. The existence of such black holes, generally called BTZ
black holes. Also, in subsequent works \cite{Strominger1998,Birmingham1998},
the three-dimensional black holes were seriously investigated, particularly
in the context of the anti-de Sitter (AdS)/conformal field theory (CFT)
correspondence (or AdS/CFT correspondence) \cite{Maldacena1998}. Indeed the
study of BTZ black holes is beneficial to comprehension gravitational
interactions in low-dimensional spacetime \cite{Witten2007}. In this regard,
some interesting works on three-dimensional black holes have been perused in
Refs. \cite{Horowitz1,Horowitz2,Import1,Horowitz3,Sfetsos,Import2,Import3}.

{Notably, when faced with the computational and conceptual challenges
of quantum gravity, one is compelled to find simple models to overcome these
significant challenges; ideally, ones that preserve some of the original
conceptual complexities while simplifying the computational effort. An
example of such a model is general relativity (GR) in $\mathbf{2+1-}$%
dimensions. Spacetime geometry in $\mathbf{2+1-}$dimensions shares many
fundamental issues with theories in $\mathbf{3+1-}$dimensions, which is an
excellent laboratory for many theoretical approaches. In addition,
fundamental physics, especially field theories in $\mathbf{2+1-}$dimensions
such as quantum hall effect, cosmic topology, parity violation, cosmic
strings, and induced masses enjoy curious properties which invite dedicated
studies \cite{54,58,59,60,61,64,65,66}. On the other hand, it has been shown
that the classical black holes (such as BTZ black holes) cannot exist in de
Sitter spacetime \cite{Emparan}. So, we consider the AdS case for gravastars
in $\mathbf{2+1-}$dimensional spacetime.}

Magueijo and Smolin \cite{MagueijoS2004} obtained the gravity's rainbow
which is the generalization of doubly special relativity for curved
spacetimes. The gravity's rainbow produces a rectification to the spacetime
that becomes substantial as soon as the particle's energy-momentum
approaches the Planck energy. So that in this formalism, connectivity and
curvature are related to the energy. In fact, in the gravity's rainbow, the
gravitational effects, in addition toas well as a possible explanation for
the absence of black holes at the LHC creating curvature in spacetime,
create different effects those are proportional to different wavelengths in
the structure of spacetime. Thus gravity's rainbow is a falsification of
spacetime defined by two arbitrary functions $L(\varepsilon )$ and $%
H(\varepsilon )$ (they are called the rainbow functions). This theory could
provide a solution for information paradox \cite{RainParaI,RainParaII}, the
existence of remains for black holes after vaporization \cite{ramnant}, as
well as a possible explanation for the absence of black holes at the LHC
\cite{LHC}. In astrophysical point of view, it has been shown that the
maximum mass of neutron stars can be more than three times the mass of Sun
\cite{NS1,NS2}. Indeed the mass of neutron stars is an increasing function
of rainbow functions \cite{NS1,NS2,NS3}. In the context of gravity, it has
been pointed out that by considering appropriate scaling for the energy
functions, the Horava-Lifshitz gravity can be related to the gravity's
rainbow \cite{HLRianbow}. In the context of black holes, various studies
focusing on the effects of gravity's rainbow on the thermodynamics of black
holes have been made in Refs. \cite{TDR1,TDR2,TDR4,TDR5,TDR6,TDR8}. Also, it
has been shown that the second law of thermodynamics and cosmic censorship
surmise are refracted owing to the rainbow effect \cite{WeakSR}. The rainbow
functions have a very considerable effect on the information flux of black
holes \cite{fluxR}. Alencar et al. \cite{Alencar}, studied the influence of
gravity's rainbow on the global Casimir effect around a static small black
hole at zero and the finite temperature. In addition, combination of
gravity's rainbow with modified theories of gravity such as $F(R)$ and $F(T)$
theories \cite{FRI,FRII,FRIII,FT}, Rastall gravity \cite{Rastall}, massive
gravity \cite{massive}, dilaton gravity \cite{dilaton}, have been also made.
{In the context of cosmology}, attention of gravity's rainbow could dispel
the big bang singularity \cite{BBangI,BBangII,BBangIII}. In addition, the
primary singularity problem \cite{KhodadiI} and stability of Einstein's
static universe in gravity's rainbow have been investigated \cite{KhodadiII}.

In order to detect a non-singular replaced to BTZ black holes, it is
necessary to explore gravastars in three-dimensional spacetime. Usmani et
al. \cite{9} have presented a new model of a gravastar admitting conformal
motion which includes a charged interior for four-dimensional spacetime.
Their exterior has been discussed by a Reissner-Nordstr\"{o}m line element
instead of Schwarzschild one. The gravastar model in higher dimensional
spacetime defined in Refs. \cite{7,9b,9c}. Later Rahaman et al. \cite{10}
have designed an uncharged gravastar in three-dimensional AdS spacetime. The
exterior region of this gravastar corresponds to the outer three-dimensional
AdS spacetime of BTZ black holes. Then, they extended the gravastar in
three-dimensional AdS spacetime by adding the electrical charge \cite{11}.
Other works on the (un)charged gravastar models with their physical features
have been studied in Refs. \cite%
{13,14,15,16,17,18,19,201,202,203,204,205,206}. Also, many gravastars have
been studied in the framework of modified theories of gravity. For example,
the gravastar solution in the $f(R,T)$ gravity model has been evaluated in
Refs. \cite{21,22,22b}. The gravastar solution in $f(G,T)$ gravity model has
been analyzed in \cite{23}. The structure of a gravastar admitting conformal
motion for a specific model of energy-momentum squared gravity has been
studied in Ref. \cite{24}. Isotropic static spherically symmetric uncharged
gravastars under the framework of braneworld gravity by using the metric
potential of Kuchowicz type have been found in Ref. \cite{Sokoliuk2022}.
Ghosh et al. \cite{Ghosh2021} extracted the gravastar solutions in Rastall
gravity. Considering $F(R,G)$ theory of gravity, the gravastar model has
been evaluated in Ref. \cite{Bhatti2020}. Debnath has discussed the geometry
of the four-dimensional charged gravastar model in Rastall-gravity's rainbow
\cite{Debnath2021}.

The basic motivation of the work is to study the gravastar in
three-dimensional AdS spacetime, both charged and uncharged states, in the
context of gravity's rainbow theory with the isotropic fluid. Also, we
research the nature of physical parameters and study the stability of
thin-shell gravastars for three-dimensional (un)charged AdS in this theory
of gravity. In this regard, the general formalism of gravity's rainbow by
using the deformation of the standard energy-momentum relation, could be
obtained
\begin{equation}
E^{2}L^{2}(\varepsilon )-p^{2}H^{2}(\varepsilon )=\mu ^{2},
\end{equation}%
where $\varepsilon =\frac{E}{E_{p}}$ . Here $p$, $E$, $\mu $ and $E_{p}=%
\sqrt{\frac{\hbar c^{5}}{G}}$ are the momentum, energy, mass of a test
particle and Planck energy, respectively. Also, $L(\varepsilon )$ and $%
H(\varepsilon )$ are called the rainbow functions. It is significant that
there are three known cases for the rainbow function;

\begin{itemize}
\item The first is related to the hard spectra caused gamma-ray bursts \cite%
{Amelino1998}, with the form
\begin{equation}
L(\varepsilon )=\frac{e^{\beta \varepsilon }-1}{\beta \varepsilon }%
~~~~\&~~~~H(\varepsilon )=1.
\end{equation}

\item The second set of rainbow functions comes from considering the
constancy of the velocity of light \cite{MagueijoS2002},
\begin{equation}
L(\varepsilon )=H(\varepsilon )=\frac{1}{1-\lambda \varepsilon }.
\end{equation}

\item The third is motivated by studying the conducted in loop quantum
gravity and noncommutative geometry as \cite{Jacob2010,Amelino2013}
\begin{equation}
L(\varepsilon )=1~~~~\&~~~~H(\varepsilon )=\sqrt{1-\eta \varepsilon ^{n}},
\end{equation}
where $\beta $, $\lambda $, $\eta $ and $n$ are constants those can be
adjusted by experiment. It is notable that at low energy levels, the rainbow
functions satisfy the following relations,
\begin{equation}
\lim_{\varepsilon \longrightarrow 0}L(\varepsilon
)=1~~~~\&~~~~\lim_{\varepsilon \longrightarrow 0}H(\varepsilon )=1.
\end{equation}
\end{itemize}

The following recipe can be used to build an energy dependent spacetime,
\begin{equation}
\widehat{g}(\varepsilon )=\eta ^{ab}e_{a}(\varepsilon )\otimes
e_{b}(\varepsilon ),  \label{rainmetric}
\end{equation}%
where $e_{0}(\varepsilon )=\frac{1}{L(\varepsilon )}\widetilde{e}_{0}$ and $%
e_{i}(\varepsilon )=\frac{1}{H(\varepsilon )}\widetilde{e}_{i}$. Also, $%
\widetilde{e}_{0}$ and $\widetilde{e}_{i}$ being the energy independent
frame fields. Considering Eq. (\ref{rainmetric}), we obtain a
three-dimensional spacetime describing the interior spacetime of a compact
object model in gravity's rainbow as
\begin{equation}
{ds}^{2}=\frac{f(r)}{L^{2}(\varepsilon )}{dt}^{2}-\frac{1}{H^{2}(\varepsilon
)}\left( \frac{{dr}^{2}}{g(r)}+r^{2}{d\theta }^{2}\right) ,  \label{metric}
\end{equation}%
where $f(r)$ and $g(r)$ are functions of $r$. Also, $L(\varepsilon )$ and $%
H(\varepsilon )$ depend on $\varepsilon =E/E_{p}$. By considering that the
metric coefficients are dependent on the energy of the test particle, the
spacetime geometry becomes energy dependent.

The structure of present work is as follows; In section \ref{II}, we deal
with the geometry of uncharged gravastar, and also we calculate the
solutions in the three regions of gravastar. Also in this section, we
analyze the physical aspects of the gravastar model parametersl. Section \ref%
{III} studies the geometry of the charged gravastar, where the solutions of
the all regions of the charged gravastar are calculated. Then we examine the
physical aspects of the parameters of charged gravastar model. Section \ref%
{IV} investigates the matching of thin shell gravastars' interior and
exterior regions for three-dimensional uncharged and charged AdS in
gravity's rainbow. In section \ref{V} we study the stability of thin shell
gravastars. Finally, some conclusions are drawn in section \ref{VI}.

%31-6-1401 at 11:20 am

\section{Three-dimensional uncharged gravastars in gravity's rainbow}

\label{II}

In this section, we obtain the solutions of field equations for gravastar in
gravity's rainbow without electromagnetic field, and analyze its geometrical
as well as physical interpretations. As it was mentioned in previous
section, there are three regions of the gravastar structured as follows; (i)
interior region $R_{1}$: $0<r<r_{1}$ with the equation of state (EOS)
follows $p=-\rho $, (ii) shell region $R_{2} $: $r_{1}<r<r_{2}$ with EOS
follows $p=\rho $, (iii) exterior region $R_{3}$: $r_{2}<r$ with EOS follows
$p=\rho =0$.

The field equation with the energy-dependent cosmological constant $\Lambda
(\varepsilon )$ and perfect fluid distribution in gravity's rainbow is as
follows
\begin{equation}
R_{\alpha \beta }-\frac{1}{2}Rg_{\alpha \beta }+\Lambda \left( \varepsilon
\right) g_{\alpha \beta }=-8\pi \left( T_{\alpha \beta }^{PF}\right) ,
\end{equation}%
where $R$ refers to the Ricci scalar, $\Lambda (\varepsilon )$ is the
cosmological constant which depends on $\varepsilon $, and $R_{\alpha \beta
} $ is the Ricci tensor. Here, we consider the matter distribution in the
interior of a compact object as a perfect fluid model given by
\begin{equation}
T_{\alpha \beta }^{PF}=(\rho +p)u_{\alpha }u_{\beta }-pg_{\alpha \beta },
\label{T}
\end{equation}%
where $p$, $\rho $, $u_{\alpha }$ are the fluid pressure, matter-energy
density and velocity three-vector of a fluid element, respectively. %
%The Einstein's field equations with a negative energy-dependent cosmological constant ($%
%\Lambda (\varepsilon )<0$) for the spacetime %
%
The Einstein's field equations with a negative energy-dependent cosmological
constant ($\Lambda (\varepsilon )<0$), for the spacetime explained by Eq. (%
\ref{metric}) with the energy-momentum tensor described in Eq. (\ref{T}),
yield (rendering $G=c=1$)%
\begin{eqnarray}
&\frac{H^{2}(\varepsilon )g^{\prime }(r)}{2r}=8\pi \rho (r)+\Lambda
(\varepsilon ),&  \label{ee1} \\
&\frac{H^{2}(\varepsilon )g(r)f^{\prime }(r)}{2rf(r)}=-8\pi p(r)+\Lambda
(\varepsilon ),&  \label{ee2} \\
&-\frac{H^{2}(\varepsilon )g(r)f^{\prime 2}(r)}{4f^{2}(r)}+\frac{%
H^{2}(\varepsilon )g^{\prime }(r)f^{\prime }(r)}{4f(r)}+\frac{%
H^{2}(\varepsilon )g(r)f^{\prime \prime }(r)}{2f(r)}=-8\pi p(r)+\Lambda
(\varepsilon ).&  \label{ee3}
\end{eqnarray}%
where the prime and double prime define the first and second derivatives
with respect to $r$, respectively. Now, the generalized
Tolman-Oppenheimer-Volkov (TOV) equation can be written as
\begin{equation}
p^{\prime }+(p+\rho )\frac{f^{\prime }}{2f}=0.  \label{tov}
\end{equation}

\subsection{Interior region of gravastar}

The interior region $R_{1}$ ($0<r<r_{1}=D$) of the gravastar follows EoS $%
p=-\rho $. We advert that this type of EoS is available in the works and is
known as a degenerate vacuum, false vacuum, or $\rho $-vacuum \cite%
{29,30,31,32} and represents a repellent pressure. Therefore by using the
result given in Eq. (\ref{tov}), we can obtain the following interior
\begin{equation}
\rho =constant=\rho _{v}.
\end{equation}

Here we define this constant as $\rho _{v}=H_{0}^{2}/2\pi $ \cite{10}, where
$H_{0}$ refers to the Hubble parameter. In other word, it can be described
as follows,
\begin{equation}
p=-\rho _{v}.
\end{equation}

Now, by using Eqs. (\ref{ee1}) and (\ref{ee2}), we get the solutions for $%
g(r)$ and $f(r)$ from the field equations as,
\begin{equation}
g(r)=f(r)=A+\frac{\left( \Lambda (\varepsilon )+4H_{0}^{2}\right) r^{2}}{%
H^{2}(\varepsilon )},  \label{g(r)}
\end{equation}%
where $A$ is an integration constant. From the above equation, it can be
seen that the obtained metric is free from any central singularity. So, the
active gravitational mass $M(r)$ can be expressed at in the following form
\begin{equation}
M(r)=\int_{0}^{D}2\pi r\rho dr=\frac{H_{0}^{2}D^{2}}{2}.
\end{equation}

Here, it is observed that for the interior region, the physical parameters
such as pressure, density and gravitational mass, do not depend on the
rainbow functions. We also note that the quantities $g(r)$ and $f(r)$ depend
on the rainbow function $H(\varepsilon )$.

\subsection{Exterior region of gravastar}

For the vacuum exterior region, EoS is given by $p=\rho =0$. The solution
corresponds to the static BTZ black hole in gravity's rainbow is written as
follows \cite{33a,33b},
\begin{equation}
f(r)=g(r)=-m_{0}-\frac{\Lambda (\varepsilon )r^{2}}{H^{2}(\varepsilon )},
\end{equation}%
where the parameter $m_{0}$ is an integration constant related to the total
mass of black hole.

\subsection{Sell of gravastar}

Here we study a thin shell including ultra-relativistic fluid of soft quanta
obeying EoS as $p=\rho $. This assumption has already been used by various
authors, which is known as a stiff fluid that refers to a Zeldovich Universe
to investigate some cosmological \cite{34, 35, 36} and astrophysical
phenomena and astrophysical events \cite{37, 38, 39}. It is not simple to
solve the field equations within the non-vacuum region $R_{2}$, i.e., within
the shell. However, we can obtain an analytic solution within the framework
of very thin shell limit, $0<g(r)\equiv h<<1$. The reason of using this very
thin shell limit is that in this limit, it can be set $h$ to be zero to the
first order. Therefore, the field equations (\ref{ee1}-\ref{ee3}), with $%
p=\rho $, can be written as,
\begin{equation}
h^{\prime }=\frac{4r\Lambda (\varepsilon )}{H^{2}(\varepsilon )}.  \label{h1}
\end{equation}

Integrating Eq. (\ref{h1}) yields
\begin{equation}
h=g(r)=B+\frac{2\Lambda (\varepsilon )r^{2}}{H^{2}(\varepsilon )}.
\end{equation}

By equating Eqs. (\ref{ee2}) and (\ref{ee3}), one gets
\begin{equation}
-\frac{1}{2}\left( {\frac{f^{\prime }}{f}}\right) ^{2}+\frac{1}{2}\left(
\frac{g^{\prime }}{g}\right) \left( \frac{f^{\prime }}{f}\right) +\frac{%
f^{\prime \prime }}{f}=\frac{f^{\prime }}{rf},
\end{equation}
therefore, the other function is
\begin{equation}
f(r)={\left( C_{1}+C_{2}\frac{H^{2}(\varepsilon )}{4\Lambda (\varepsilon )}%
\sqrt{B+\frac{2\Lambda (\varepsilon )r^{2}}{H^{2}(\varepsilon )}}\right) }%
^{2},
\end{equation}%
where $B$, and $C_{1}$ are integration constants. It is notable that $C_{2}$
is a constant with a dimension as (length)$^{-2}$ that is considered for the
sake of having a dimensionless term. Also, from the conservation (Eq. (\ref%
{tov})), and using EOS $p=\rho $, one can obtain
\begin{equation}
p=\rho =p_{0}{\left( C_{1}+C_{2}\frac{H^{2}(\varepsilon )}{4\Lambda
(\varepsilon )}\sqrt{B+\frac{2\Lambda (\varepsilon )r^{2}}{H^{2}(\varepsilon
)}}\right) }^{-2},
\end{equation}%
where $p_{0}$ being an integration constant.

\subsection{Physical parameters}

\label{the physical parameters}

\subsubsection{Proper Length}

We note that the radius of inter boundary of shell is $r_{1}=D$ and the
radius of the exterior boundary of shell is $r_{2}=D+\delta $, where $\delta
$ is the proper thickness of shell which is considered to be very small
(i.e., $\delta <<1$). The proper length of shell is described as $%
l=\int_{D}^{D+\delta }\sqrt{\frac{1}{g(r)}}dr$ \cite{1}. Applying the
mentioned definition, we extract the proper length in gravity's rainbow,
which is given by
\begin{equation}
l=\int_{D}^{D+\delta }\sqrt{\frac{1}{H^{2}(\varepsilon )g(r)}}dr.
\end{equation}

Since $g(r)$ is complicated in the shell region, it is not simple to obtain
the analytical form of above integral. Here we assume $\sqrt{\frac{1}{g(r)}}=%
\frac{dF(r)}{dr}$, so from the above integral can be written
\begin{eqnarray}
l &=&\frac{1}{H(\varepsilon )}\int_{D}^{D+\delta }\frac{dF(r)}{dr}dr=\frac{%
F(D+\delta )-F(D)}{H(\varepsilon )}  \notag \\
&&  \notag \\
&\approx &\frac{\delta }{H(\varepsilon )}\frac{dF(r)}{dr}\mid _{_{r=D}}=%
\frac{\delta }{H(\varepsilon )}\sqrt{\frac{1}{g(r)}}\mid _{_{r=D}}.
\end{eqnarray}

Since $\delta <<1$ ($O(\delta ^{2})\approx 0$), according to the above
manipulation, it can be considered only the first-order term of $\delta $.
So for this approximation, the proper length will be equal to
\begin{equation}
l\approx \frac{\delta }{H(\varepsilon )}\sqrt{\frac{1}{B+\frac{2\Lambda
\left( \varepsilon \right) D^{2}}{H^{2}(\varepsilon )}}}.
\end{equation}

From the above result, we understand that the proper length of thin shell of
the gravastar is proportional to the thickness ($\delta $) of shell. Also it
is observed that the proper length of thin shell depends on the rainbow
function $H(\varepsilon )$.

\subsubsection{Energy}

It is assumed that the energy inside the gravastar has dark energy, which
actually the repulsive force from the interior comes from this energy. The
energy content within the shell region of the gravastar described as $%
E=\int_{D}^{D+\delta }{2\pi }r\rho dr$ \cite{1}. So, we obtain the energy
content within the thin shell in gravity's rainbow, which is given as
\begin{equation}
E=\frac{1}{H^{2}(\varepsilon )}\int_{D}^{D+\delta }2\pi r\rho dr.
\end{equation}

Similarly, as done above, by expanding $F(D+\delta )$ binomially about $D$
and taking the first order of $\delta $, we obtain the following relation,
\begin{equation}
E\approx \frac{2\pi p_{0}\delta D}{H^{2}(\varepsilon ){\left( C_{1}+C_{2}%
\frac{H^{2}(\varepsilon )}{4\Lambda (\varepsilon )}\sqrt{B+\frac{2\Lambda
(\varepsilon )D^{2}}{H^{2}(\varepsilon )}}\right) }^{2}}.
\end{equation}

%Another physical parameter that we want to evaluate is related to entropy.

\subsubsection{Entropy}

%{Entropy, in fact, is the disorderliness inside the body of a gravastar.}
Mazur and Mottola \cite{1,2} showed that the entropy density is zero in the
interior region $R_{1}$ of the gravastar. But, the entropy within the thin
shell in gravity's rainbow can be expressed by
\begin{equation}
S=2\pi \int_{D}^{D+\delta }\sqrt{\frac{1}{H^{2}(\varepsilon )g(r)}}s(r)rdr,
\end{equation}%
where $s(r)$, the entropy density for the local temperature $T(r)$, is given
by
\begin{equation}  \label{s1}
s(r)=\frac{\alpha ^{2}k_{B}^{2}T(r)}{4\pi \hbar ^{2}}=\frac{\alpha k_{B}}{%
\hbar }\sqrt{\frac{p(r)}{2\pi }}.
\end{equation}

In above equation, $k_{B}$ is Boltzmann constant, $\hbar $ is Reduced Planck
constant and $\alpha ^{2}$ is a dimensionless constant. The entropy within
the thin shell can be written as
\begin{equation}
S=\frac{\alpha k_{B}\sqrt{2\pi p_{0}}}{\hbar H(\varepsilon )}%
\int_{D}^{D+\delta }\frac{rdr}{\sqrt{B+\frac{2\Lambda (\varepsilon )r^{2}}{%
H^{2}(\varepsilon )}}\left( C_{1}+C_{2}\frac{H^{2}(\varepsilon )}{4\Lambda
(\varepsilon )}\sqrt{B+\frac{2\Lambda (\varepsilon )r^{2}}{H^{2}(\varepsilon
)}}\right) }.
\end{equation}

Similarly, by expanding $F(D+\delta )$ binomially about $D$, it is obtained
\begin{equation}
S\approx \frac{D\delta \alpha k_{B}\sqrt{2\pi p_{0}}}{\hbar H(\varepsilon )%
\sqrt{B+\frac{2\Lambda (\varepsilon )D^{2}}{H^{2}(\varepsilon )}}\left(
C_{1}+C_{2}\frac{H^{2}(\varepsilon )}{4\Lambda (\varepsilon )}\sqrt{B+\frac{%
2\Lambda (\varepsilon )D^{2}}{H^{2}(\varepsilon )}}\right) }.
\end{equation}

\begin{figure}[tbp]
\centering
\includegraphics[scale=0.3]{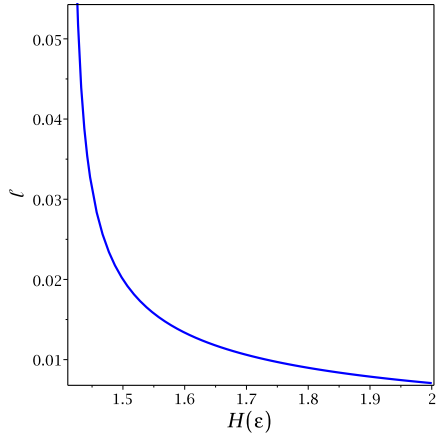}
\caption{The proper length ($l$) vs the rainbow function ($H(\protect%
\varepsilon )$). We have chosen $\protect\delta =0.01$, $D=1$, $\Lambda (%
\protect\varepsilon )=-1$, and $B=1$.}
\label{plot l}
\end{figure}

\begin{figure}[tbp]
\centering
\includegraphics[scale=0.3]{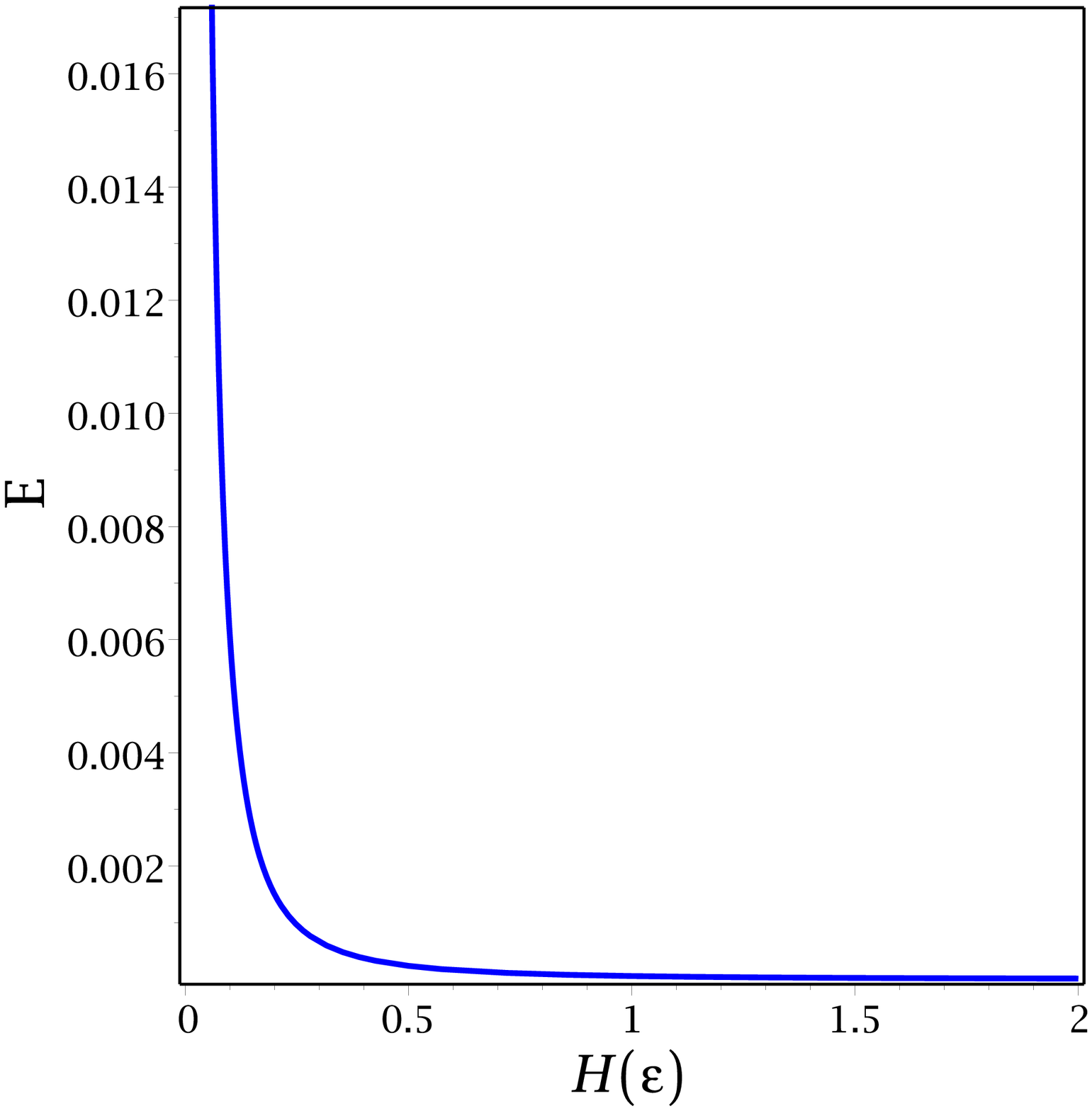}
\caption{The energy $E$ vs the rainbow function ($H(\protect\varepsilon )$).
We have chosen $\protect\delta =0.01$, $D=1$, $\Lambda (\protect\varepsilon %
)=-1$, $C_{1}=10$, $C_{2}=1$, $p_{0}=1$, and $B=1$.}
\label{plot E}
\end{figure}

\begin{figure}[tbp]
\centering
\includegraphics[scale=0.3]{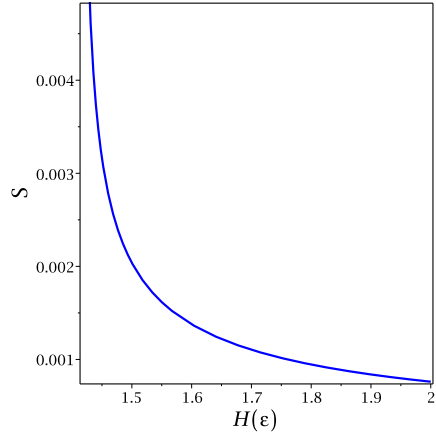}
\caption{The entropy $S$ vs the rainbow function ($H(\protect\varepsilon )$%
). We have chosen $\protect\delta =0.01$, $D=1$, $\Lambda (\protect%
\varepsilon )=-1$, $C_{1}=10$, $C_{2}=1$, $p_{0}=1$, and $B=1$.}
\label{s}
\end{figure}

Here we have plotted the physical parameters, the proper length $l$, the
energy $E$, and the entropy $S$ versus the rainbow function ($H(\varepsilon
) $) in Figs. \ref{plot l}-\ref{s}, respectively. From these figures, we see
that all three parameters of the shell of gravastar decrease versus the
rainbow function $H(\varepsilon )$.

\section{Three-dimensional Charged AdS Gravastars in Gravity's Rainbow}

\label{III}

\label{section.charged gravastars in gravity's rainbow}

The Einstein-Hilbert action coupled to the energy-dependent cosmological
constant and electromagnetism is given by
\begin{equation}
I=\int {d}^{3}{x}\sqrt{-g}\left( \frac{R-2\Lambda \left( \varepsilon \right)
}{16\pi }-\frac{1}{4}F_{\mu \nu }F^{\mu \nu }+L_{m}\right),
\end{equation}%
where $L_{m}$ is Lagrangian for the matter. It is assumed that the fluid
source consists of natural matter and an electromagnetic field. The
Einstein-Maxwell equations with the energy-dependent cosmological constant $%
\Lambda (\varepsilon )$ for a charged perfect fluid distribution is as
follows \cite{CataldoC2006},
\begin{equation}
R_{\alpha \beta }-\frac{1}{2}Rg_{\alpha \beta }+\Lambda \left( \varepsilon
\right) g_{\alpha \beta }=-8\pi \left( T_{\alpha \beta }^{PF}+T_{\alpha
\beta }^{EM}\right),
\end{equation}%
where the energy-momentum tensor $T_{\alpha \beta }^{EM}$ for normal matter
is given by
\begin{equation}
T_{\alpha \beta }^{PF}=(\rho +p)u_{\alpha }u_{\beta }-pg_{\alpha \beta },
\end{equation}
here $\rho $, $p$, $u_{i}$ are respectively, matter-energy density, fluid
pressure and velocity three-vector of a fluid element. The energy-momentum
tensor for electromagnetic field is described by,
\begin{equation}
T_{\alpha \beta }^{EM}=-\frac{1}{4}\left( F_{\gamma }^{\alpha }F_{\beta
\gamma }-\frac{1}{4}g_{\alpha \beta }F_{\gamma \kappa }F^{\gamma \kappa
}\right),
\end{equation}%
where $F_{\alpha \beta }$ is the electromagnetic field and depends on
current three-vector,
\begin{equation}
J^{\gamma }=\sigma (r)u^{\gamma },
\end{equation}%
as
\begin{equation}
F_{;\beta }^{\alpha \beta }=-4\pi J^{\alpha }.
\end{equation}%
where $\sigma (r)$ is the proper charge density. The three-velocity is
considered as $u_{\alpha }=\delta _{\alpha }^{t}$, and eventually, the
electromagnetic field tensor can be described as
\begin{equation}
F_{\alpha \beta }=E(r)\left( \delta _{\alpha }^{t}\delta _{\beta
}^{r}-\delta _{\alpha }^{r}\delta _{\beta }^{t}\right) ,
\end{equation}%
where $E(r)$ is the electric field.

For a the charged fluid source with density $\rho (r)$, pressure $p(r)$ and
electric field $E(r)$, the Einstein-Maxwell equations in the gravity's
rainbow can be described as follows,
\begin{eqnarray}
&\frac{H^{2}(\varepsilon )g^{\prime }(r)}{2r} =8\pi \rho (r)+\Lambda
(\varepsilon )+E^{2}(r),&  \label{eq11} \\
&\frac{H^{2}(\varepsilon )g(r)f^{\prime }(r)}{2rf(r)} =-8\pi p(r)+\Lambda
(\varepsilon )+E^{2}(r),&  \label{eq12} \\
&-\frac{H^{2}(\varepsilon )g(r)f^{\prime 2}(r)}{4f^{2}(r)}+\frac{%
H^{2}(\varepsilon )g^{\prime }(r)f^{\prime }(r)}{4f(r)}+\frac{%
H^{2}(\varepsilon )g(r)f^{\prime \prime }(r)}{2f(r)} =-8\pi p(r)+\Lambda
(\varepsilon )-E^{2}(r).&  \label{eq13}
\end{eqnarray}

Now, for a charged fluid distribution, the generalized TOV equation can be
expressed as
\begin{equation}
p^{\prime }+(p+\rho )\frac{f^{\prime }}{2f}=\frac{1}{8\pi r^{2}}%
(r^{2}E^{2})^{\prime },  \label{eq14}
\end{equation}%
where $E\equiv E(r)$, and the electric field $E$ is as follows
\begin{equation}
E(r)=\frac{4\pi }{r}\int_{0}^{r}\frac{x\sigma (x)dx}{H(\varepsilon )\sqrt{%
g(r)}}=\frac{q(r)}{r}.  \label{eq15}
\end{equation}

The parameter $\frac{\sigma (x)}{H(\varepsilon )\sqrt{g(r)}}$, which is
inside the above integral, is considered as the volume charge density. It
should be noted that the charge density volume is a polynomial function of $%
r $. Therefore, we use the condition
\begin{equation}
\frac{\sigma (x)}{H(\varepsilon )\sqrt{g(r)}}=\sigma _{0}r^{m},  \label{eq16}
\end{equation}%
were $m$ is an arbitrary constant introduced as a polynomial index and the
constant $\sigma _{0}$ is related to as the central charge density. By using
the result in Eq. (\ref{eq16}), one obtains from Eq. (\ref{eq15}),
\begin{eqnarray}
E(r) &=&\frac{4\pi \sigma _{0}}{m+2}r^{m+1},  \label{eq17} \\
&&  \notag \\
q(r) &=&\frac{4\pi \sigma _{0}}{m+2}r^{m+2}.
\end{eqnarray}

Now, we analyze the solutions of field equations for charged gravastar
(here, charge is created by electromagnetic field) and discuss its geometry
as well as physical interpretations.

\subsection{Interior region of charged gravastar}

In search for interior solution which is free of any mass-singularity at the
origin, our configuration is supported by an interior region $R_{1}$ ($%
0<r<r_{1}=D$) of the charged gravastar with equation of state $p=-\rho $.
Hence by using the result given in Eqs. (\ref{eq11}) and (\ref{eq14}), we
obtain the following interior,
\begin{eqnarray}
\rho (r) &=&-p(r)=\frac{2\pi \sigma _{0}^{2}\left( D^{2m+2}-r^{2m+2}\right)
}{(m+1)(m+2)},  \label{eq19} \\
&&  \notag \\
g(r) &=&C_{0}+\frac{\left[ 16\pi \int_{0}^{r}x\rho
(x)dx+2\int_{0}^{r}xE^{2}(x)dx+\Lambda \left( \varepsilon \right) r^{2}%
\right] }{H^{2}(\varepsilon )}, \\
&&  \notag \\
f(r) &=&g(r)=C_{0}+\frac{\left[ \frac{16\pi ^{2}\sigma _{0}^{2}r^{2}D^{2m+2}%
}{(m+1)(m+2)}-\frac{16\pi ^{2}\sigma _{0}^{2}r^{2m+4}}{(m+1)({m+2})^{3}}%
+\Lambda \left( \varepsilon \right) r^{2}\right] }{H^{2}(\varepsilon )},
\label{eq21}
\end{eqnarray}%
where $C_{0}$ is an integration constant. We note from Eq. (\ref{eq21}) that
$C_{0}$ is a nonzero integration constant. We get the charge density for the
electric field in the following form,
\begin{equation}
\sigma (r)=\sigma _{0}r^{m}\left( C_{0}+\frac{1}{H^{2}(\varepsilon )}\left[
\frac{16\pi ^{2}\sigma _{0}^{2}r^{2}D^{2m+2}}{(m+1)(m+2)}-\frac{16\pi
^{2}\sigma _{0}^{2}r^{2m+4}}{(m+1)({m+2})^{3}}+\Lambda (\varepsilon ) r^{2}%
\right] \right) ^{1/2}.
\end{equation}

We can obtain the active gravitational mass of the interior region of the
charged gravastar as follows
\begin{equation}
M(r)=\frac{1}{H^{2}(\varepsilon )}\int_{0}^{D}2\pi r\left( \rho +\frac{E^{2}%
}{8\pi }\right) dr=\frac{(2m^{2}+8m+6)\pi ^{2}\sigma _{0}^{2}D^{2m+4}}{%
H^{2}(\varepsilon )(m+2)^{3}(m+1)} .  \label{M}
\end{equation}

A characteristic feature of a stellar compact object is that the internal
gravitational mass and radius of the star system are directly proportional
to each other, which is shown by Eq. (\ref{M}). Here, it can be also
observed via Eq. (\ref{eq19}) that the physical parameters such as pressure
and density are dependent on the charge. However, in this connection, we
note an interesting point; the mentioned physical parameters are not
dependent on the energy-dependent cosmological constant $\Lambda \left(
\varepsilon \right) $ and rainbow functions for the interior region. We also
observe that the quantities $g(r)$, $f(r)$, $M(r)$, and $\sigma (r)$ depend
not only on the rainbow function $H(\varepsilon )$, but also on the charge.

\subsection{Exterior region of charged gravastar}

In order to discuss the exterior region of charged gravastar whose EoS is
given by $p=\rho =0$, we take the solution corresponds to a static charged
BTZ black hole in the gravity's rainbow described as follows \cite{39a},
\begin{equation}
f(r)=g(r)=-m_{0}-\frac{\Lambda \left( \varepsilon \right) r^{2}}{%
H^{2}(\varepsilon )}-\frac{2q^{2}L^{2}(\varepsilon )}{l^{2}}\ln \left( \frac{%
r}{l}\right),
\end{equation}%
where the parameter $q$ is an integration constant referred to the total
charge of the black hole. Furthermore, $l$ is a constant with the length
dimension that is considered for the sake of having a dimensionless
logarithmic argument. For $q=0$, the above metric reaches to the uncharged
BTZ black hole in gravity's rainbow.

\subsection{Sell of charged gravastar}

In the shell region, it is assumed that the thin shell region contains stiff
perfect fluid which follows EoS $p=\rho $. For this EoS, it is complex to
get to the answer the solution from the field equations. Similar to the
uncharged case, we shall consider the limit $0<g(r)\equiv h<<1$ in the thin
shell to obtain the analytical solution within the thin shell. Assuming this
approximation ($h$ to be zero) with $p=\rho $, the field equations (\ref%
{eq11})-(\ref{eq13}), can be recast in the following forms,
\begin{eqnarray}
h^{\prime } &=&\frac{4r(\Lambda \left( \varepsilon \right) +E^{2})}{%
H^{2}(\varepsilon )},  \label{eq27} \\
&&  \notag \\
\frac{H^{2}(\varepsilon )h^{\prime }f^{\prime }(r)}{4f(r)} &=&-2E^{2}.
\label{eq28}
\end{eqnarray}

By integrating Eq. (\ref{eq27}), we get
\begin{equation}
g(r)=A_{0}+\frac{2\Lambda \left( \varepsilon \right) r^{2}}{%
H^{2}(\varepsilon )}+\frac{32\pi ^{2}\sigma _{0}^{2}r^{2m+4}}{%
H^{2}(\varepsilon )\left( m+2\right) ^{3}},  \label{eq29}
\end{equation}%
where $A_{0}$ is an integrating constant. Using this value in Eq. (\ref{eq28}%
), it is obtained another metric coefficient as,
\begin{equation}
f(r)=F_{0}\left( {{\Lambda \left( \varepsilon \right) +\frac{16\pi
^{2}\sigma _{0}^{2}r^{2m+2}}{{(m+2)}^{2}}}}\right) ^{\frac{-1}{m+1}},
\end{equation}%
where $F_{0}$ is an integration constant. Also, using the generalized TOV
equation (\ref{eq14}) one can get
\begin{equation}  \label{eq57}
8\pi p(r)=8\pi \rho (r)=\left( \frac{m+2}{m}\right) \left( \Lambda
(\varepsilon )+\frac{16\pi ^{2}\sigma _{0}^{2}r^{2m+2}}{(m+2)^{2}}\right),
\end{equation}
the equation (\ref{eq57}) shows that the energy-dependent cosmological
constant has a important contribution to the pressure and density parameters
of the shell of charged gravastar in an additive manner.

\subsection{Physical parameters of charged gravastar}

\label{the physical parameters charged gravastar}

\subsubsection{Proper Length}

%According to this fact that
The radius of inner and outer surfaces of the shell of gravastar
%\textquotesingle s
are $r_{1}=D$ and $r_{2}=D+\delta $, respectively. Similar to the previous
case (uncharged case), and by assuming $\sqrt{\frac{1}{g(r)}}=\frac{dF(r)}{dr%
}$, the proper thickness between two boundaries can be described as,
\begin{eqnarray}
l &=&\frac{1}{H(\varepsilon )}\int_{D}^{D+\delta }\frac{dF(r)}{dr}dr=\frac{%
F(D+\delta )-F(D)}{H(\varepsilon )}  \notag \\
&&  \notag \\
&\approx &\frac{\delta }{H(\varepsilon )}\frac{dF(r)}{dr}\mid _{_{r=D}}=%
\frac{\delta }{H(\varepsilon )}\sqrt{\frac{1}{g(r)}}\mid _{_{r=D}}.
\end{eqnarray}

As we know $\delta <<1$, so $O(\delta ^{2})\approx 0$. According to this
approximation, the proper length is given by,
\begin{equation}
l\approx \frac{\delta }{H(\varepsilon )}\left( A_{0}+\frac{2\Lambda \left(
\varepsilon \right) D^{2}}{H^{2}(\varepsilon )}+\frac{32\pi ^{2}\sigma
_{0}^{2}D^{2m+4}}{H^{2}(\varepsilon ){(m+2)}^{3}}\right) ^{-1/2}.
\end{equation}

\begin{figure}[tbp]
\centering
\includegraphics[scale=0.3]{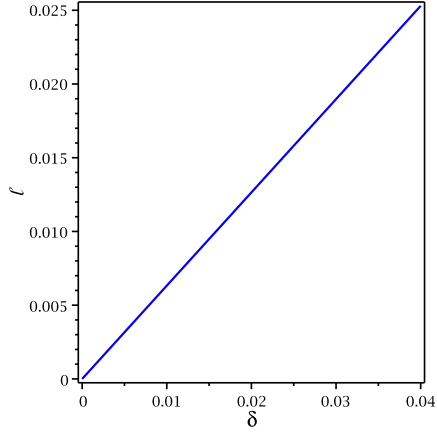}
\caption{The proper length $l$ vs thickness $\protect\delta $. We have
chosen $\protect\sigma _{0}=4$, $D=1$, $A_{0}=1$, $\Lambda (\protect%
\varepsilon )=-1$. $m=2$ and $H(\protect\varepsilon )=1.1$.}
\label{peroper.e}
\end{figure}

\begin{figure}[tbp]
\centering
\includegraphics[scale=0.3]{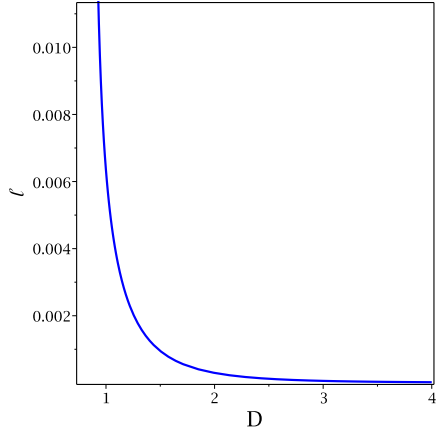}
\caption{The proper length $l$ vs radius $D$. We have chosen $\protect\sigma %
_{0}=4$, $\protect\delta =0.01$, $A_{0}=1$, $\Lambda (\protect\varepsilon %
)=-1$, $m=2$ and $H(\protect\varepsilon )=1.1$.}
\label{peroperD}
\end{figure}

From the above result, we understand that the proper length of the thin
shell of the charged gravastar is proportional to the shell's thickness $%
\delta $. Also it is observed that the proper length of thin shell of the
charged gravastar depends on the electric field $\sigma _{0}$, the radius $D$%
, the rainbow function $H(\varepsilon )$, the energy-dependent cosmological
constant $\Lambda (\varepsilon )$, and also $m$. The proper length $l$ have
been plotted versus thickness $\delta $ and radius $D$ in Figs. \ref%
{peroper.e} and \ref{peroperD}, respectively. In Fig. \ref{peroper.e}, it
can be seen a linear relationship between proper length and thickness of the
shell. Also from Fig. \ref{peroperD}, we see that the proper length $l$
decreases as radius $D$ increases.

\subsubsection{Energy}

The energy content within the shell region of the charged gravastar is
defined in the following form \cite{11},
\begin{equation}
E=\int_{D}^{D+\delta }2\pi (\rho +\frac{E^{2}}{8\pi })rdr.  \label{E}
\end{equation}

We use Eq. (\ref{E}), to calculate the energy content within the thin shell
region of the charged gravastar in gravity's rainbow as follows,
\begin{eqnarray}
E &=&\frac{2\pi \int_{D}^{D+\delta }\left( \rho +\frac{E^{2}}{8\pi }\right)
rdr}{H^{2}\left( \varepsilon \right) }  \notag \\
&&  \notag \\
&=&\frac{(m+2)\Lambda \left( \varepsilon \right) \left[ {(D+\delta )}%
^{2}-D^{2}\right] }{8mH^{2}\left( \varepsilon \right) }+\frac{16\pi
^{2}\sigma _{0}^{2}(m+1)\left[ {(D+\delta )}^{2m+4}-D^{2m+4}\right] }{4m{%
(m+2)}^{3}}.
\end{eqnarray}

According to this fact that the thickness $\delta $ of shell is very small, $%
\delta <<1$, it can be expanded binomially about $D$. Taking the first order
of $\delta $, we obtain,
\begin{equation}
E\approx \frac{\delta }{H^{2}(\varepsilon )}\left[ \frac{(m+2)\Lambda \left(
\varepsilon \right) D}{4m}+\frac{16\pi ^{2}\sigma _{0}^{2}(m+1)D^{2m+3}}{2m{%
(m+2)}^{2}}\right] .
\end{equation}

\begin{figure}[tbp]
\centering
\includegraphics[scale=0.3]{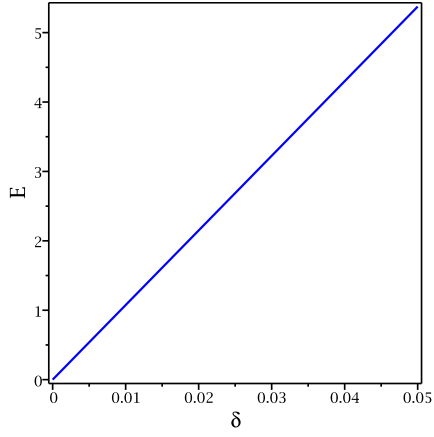}
\caption{The energy $E$ vs thickness $\protect\delta $. We have chosen $%
\protect\sigma _{0}=4$, $D=1$, $A_{0}=1$, $\Lambda (\protect\varepsilon )=-1$%
, $m=2$ and $H(\protect\varepsilon )=1.1$.}
\label{Energy.e}
\end{figure}

\begin{figure}[tbp]
\centering
\includegraphics[scale=0.3]{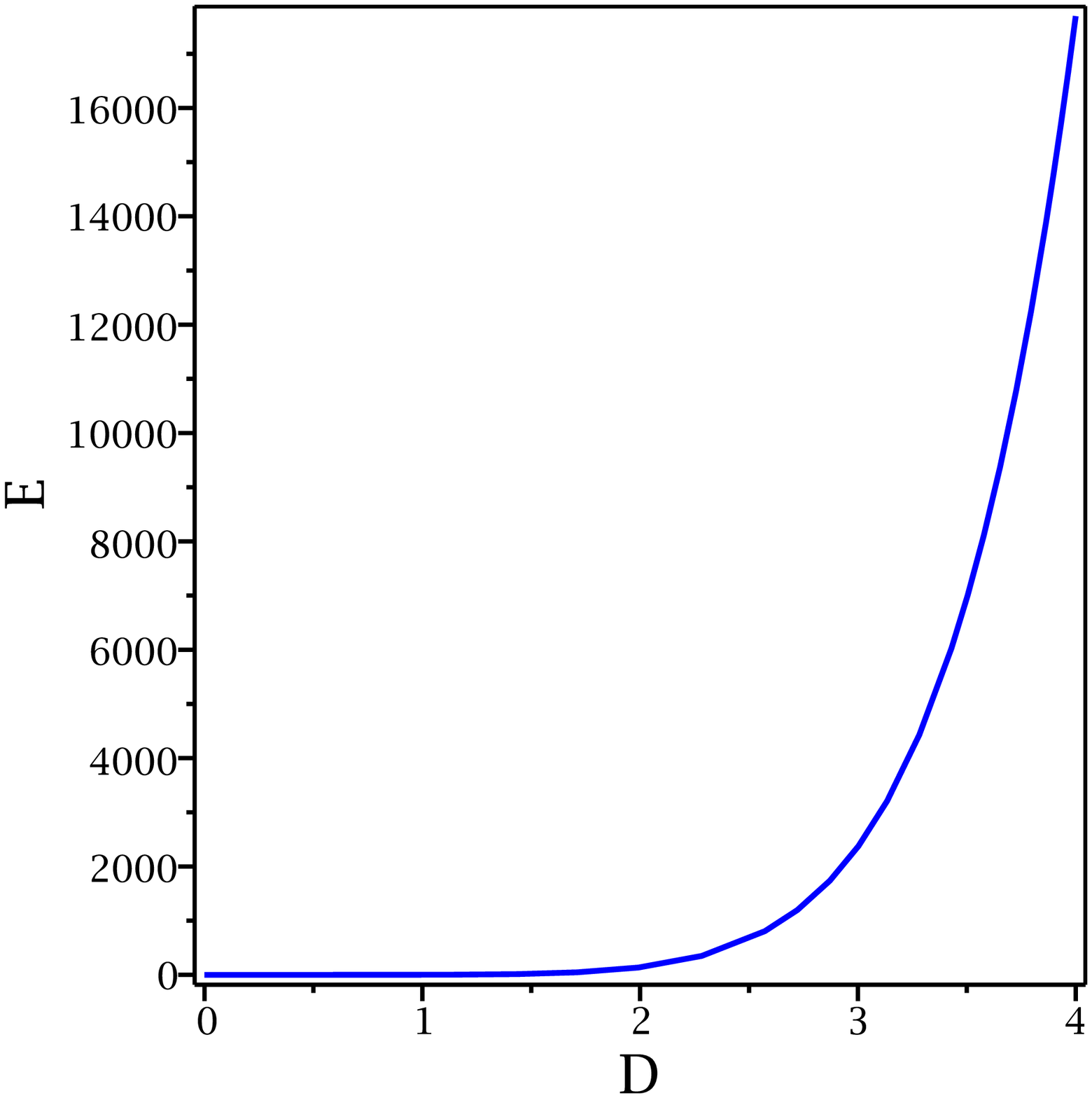}
\caption{The energy $E$ vs radius $D$. We have chosen $\protect\sigma _{0}=1$%
, $\protect\delta =0.01$, $A_{0}=1$, $\Lambda (\protect\varepsilon )=-1$, $%
m=2$ and $H(\protect\varepsilon )=1.1$.}
\label{EnergyD}
\end{figure}

We observe that the energy content in the shell region is proportional to
the thickness $\delta $ of the shell. In addition, we see that the energy of
charged gravastar depends on the electric field $\sigma _{0}$ of gravastar,
the radius $D$, the rainbow function $H(\varepsilon )$, the energy-dependent
cosmological constant $\Lambda (\varepsilon )$, and also $m$. We have
plotted the energy content $E$ in the shell versus thickness $\delta $ and
radius $D$ in Figs. \ref{Energy.e} and \ref{EnergyD}, respectively. As one
can see the energy content $E$ in the shell increases with the thickness $%
\delta $ of shell of the charged gravastar (Fig. \ref{Energy.e}) as well as
the radius $D$ (Fig. \ref{EnergyD}).

\subsubsection{Entropy}

\begin{figure}[tbp]
\centering
\includegraphics[scale=0.3]{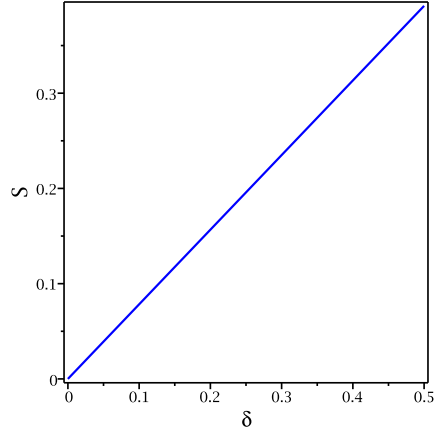}
\caption{The entropy $S$ vs thickness $\protect\delta $. We have chosen $%
\protect\sigma _{0}=4$, $D=1$, $A_{0}=1$, $\Lambda (\protect\varepsilon )=-1$%
, $m=2$ and $H(\protect\varepsilon )=1.1$.}
\label{Entropy.e}
\end{figure}

\begin{figure}[tbp]
\centering
\includegraphics[scale=0.3]{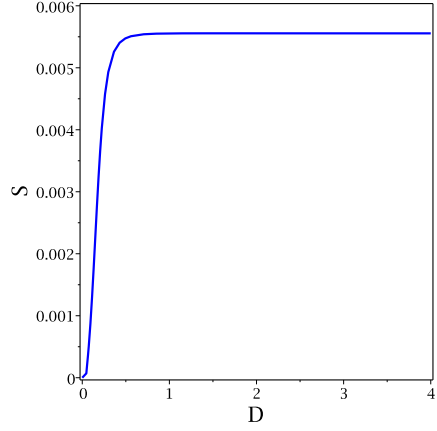}
\caption{The entropy $S$ vs radius $D$. We have chosen $\protect\sigma %
_{0}=4 $, $\protect\delta =0.01$, $A_{0}=1$, $\Lambda (\protect\varepsilon %
)=-1$, $m=1$ and $H(\protect\varepsilon )=1.1$.}
\label{Entropy-D}
\end{figure}

We obtain the entropy within the thin shell of charged gravastar in
three-dimensional spacetime as
\begin{equation}
S=2\pi \int_{D}^{D+\delta }\sqrt{\frac{1}{H^{2}(\varepsilon )g(r)}}rs(r)dr,
\end{equation}%
where $s(r)$ is the entropy density, which is given by Eq. (\ref{s1}). So
for the entropy within the thin shell, we can write
\begin{equation}
S=\frac{1}{2H(\varepsilon )}\int_{D}^{D+\delta }\frac{\alpha k_{B}}{\hbar }%
\sqrt{\frac{m+2}{m}\left[ \Lambda \left( \varepsilon \right) r^{2}+\frac{%
16\pi ^{2}\sigma _{0}^{2}r^{2m+4}}{{(m+2)}^{2}}\right] }\mathcal{A}%
_{1}^{-1/2}dr.
\end{equation}%
where $\mathcal{A}_{1}=A_{0}+\frac{2\Lambda \left( \varepsilon \right) r^{2}%
}{H^{2}(\varepsilon )}+\frac{32\pi ^{2}\sigma _{0}^{2}D^{2m+4}}{%
H^{2}(\varepsilon ){(m+2)}^{3}}$. To extract the entropy in the following
form, we use the approximation $\delta <<1$, and expand the above integrand
about $D$ and retain terms up to the first order of $\delta $,
\begin{equation}  \label{s2}
S\approx \frac{\delta \alpha k_{B}}{2H(\varepsilon )\hbar }\sqrt{\frac{m+2}{m%
}\left[ \Lambda \left( \varepsilon \right) D^{2}+\frac{16\pi ^{2}\sigma
_{0}^{2}D^{2m+4}}{{(m+2)}^{2}}\right] }\mathcal{A}_{2}^{-1/2}.
\end{equation}%
where $\mathcal{A}_{2}=A_{0}+\frac{2\Lambda \left( \varepsilon \right) D^{2}%
}{H^{2}(\varepsilon )}+\frac{32\pi ^{2}\sigma _{0}^{2}D^{2m+4}}{%
H^{2}(\varepsilon ){(m+2)}^{3}}$. Eq. (\ref{s2}) reveals that the entropy in
the shell region of the charged gravastar in three-dimensional
energy-dependent spacetime is proportional to the thickness $\delta $ of the
shell, the electric field $\sigma _{0}$, the radius $D$, the rainbow
function $H(\varepsilon )$, the energy-dependent cosmological constant $%
\Lambda (\varepsilon )$, and $m$. We have drawn the entropy $S$ within the
shell versus the thickness $\delta $ and radius $D$ in Figs. \ref{Entropy.e}
and \ref{Entropy-D}, respectively. The entropy $S$ increases when the
thickness $\delta $ and the radius $D$ increase.

%%%%%%%%%%%%%%%%%%%%%%%%%%%
%
%  3-7-1401 at 12:38
%
%%%%%%%%%%%%%%%%%%%%%%%%%%%

\section{Junction Condition between interior and exterior regions}

\label{IV}

\label{Junction Condition}

According to this fact that gravastar consists of three regions, interior
region, thin shell region, and exterior region, using the Darmois-Israel
formalism \cite{40,41,42}, one can study the matching between the surfaces
of the interior and exterior regions. For this purpose, we define $\Sigma$
as the junction surface located at $r=D$. In the gravity's rainbow, the
metric \ref{metric} considered on the junction surface. It appears that the
metric coefficients at the junction surface $\Sigma$ are continuous, but
their derivatives might not be continuous at $\Sigma$. Using the extrinsic
curvature of $\Sigma$ at $r=D$, in the joining surface $S$, it is possible
to determine the surface stress-energy and surface tension of the junction
surface $S$.

\begin{figure}[tbp]
\centering
\includegraphics[scale=0.3]{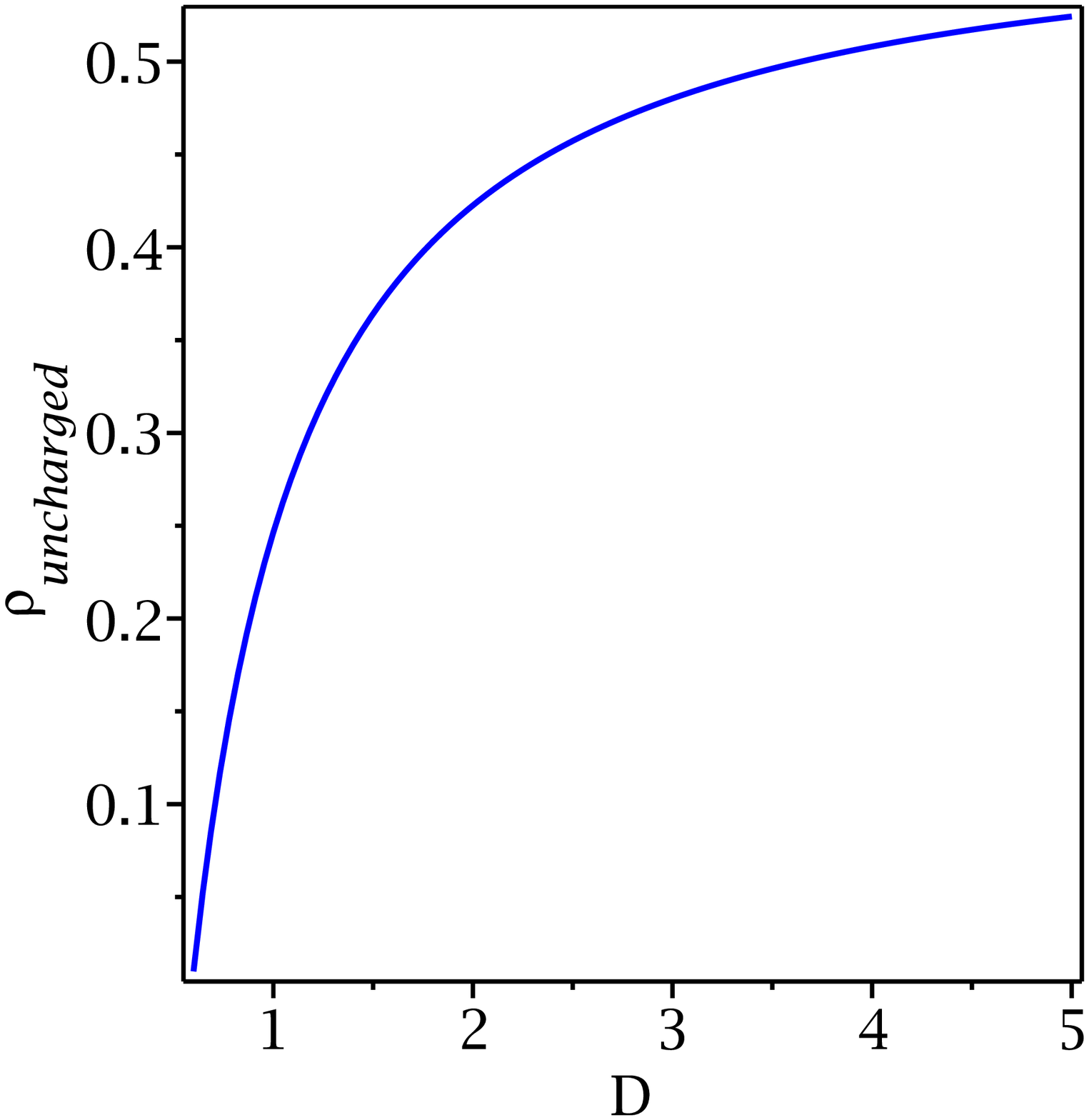}
\caption{The line energy density $\protect\varrho _{uncharged}$ vs radius $D$%
. We have chosen $m_{0}=-80$, $H_{0}=70$, $A=1$, $\Lambda (\protect%
\varepsilon )=-1$, and $H(\protect\varepsilon )=1.1$.}
\label{rhounch}
\end{figure}

\begin{figure}[tbp]
\centering
\includegraphics[scale=0.3]{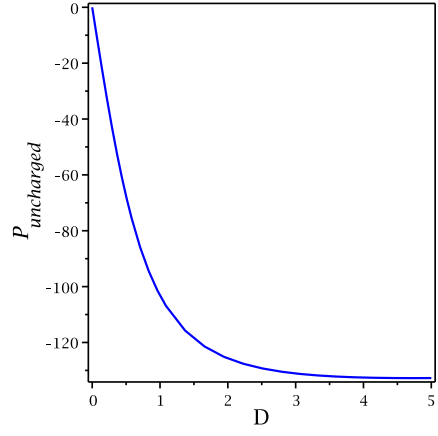}
\caption{The line pressure $P_{uncharged}$ vs radius $D$. We have chosen $%
m_{0}=-80$, $H_{0}=70$, $A=1$, $\Lambda (\protect\varepsilon )=-1$, and $H(%
\protect\varepsilon )=1.1$. }
\label{Punch}
\end{figure}

\begin{figure}[tbp]
\centering
\includegraphics[scale=0.3]{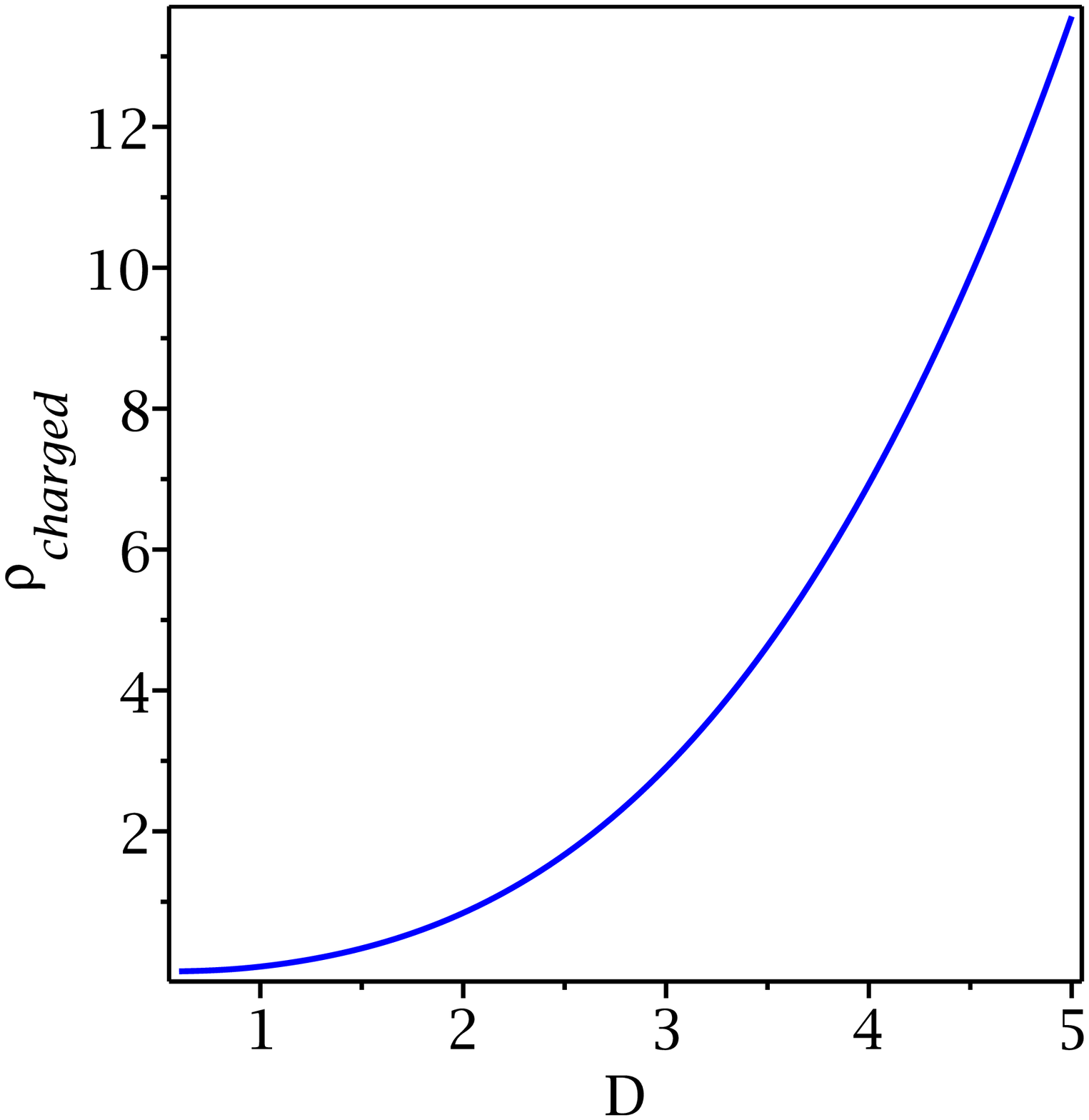}
\caption{The line energy density $\protect\varrho _{charged}$ vs radius $D$.
We have chosen $m_{0}=-80$, $L(\protect\varepsilon )=1.1$, $H(\protect%
\varepsilon )=1.1$, $\protect\sigma _{0}=4$, $C_{0}=1$, $\Lambda (\protect%
\varepsilon )=-1$, $m=1$, $q=1$, and $l=1$.}
\label{rhoch}
\end{figure}

\begin{figure}[tbp]
\centering
\includegraphics[scale=0.3]{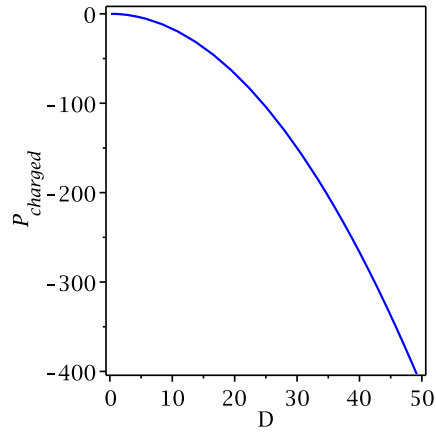}
\caption{The line pressure $P_{charged}$ vs radius $D$. We have chosen $%
m_{0}=-80$, $L(\protect\varepsilon )=1.1$, $H(\protect\varepsilon )=1.1$, $%
\protect\sigma _{0}=4$, $C_{0}=1$, $\Lambda (\protect\varepsilon )=-1$, $m=1$%
, $q=1$, and $l=1$.}
\label{Pch}
\end{figure}

{According to the Lanczos equation \cite{52}, stress-energy surface $%
S_{ij}$ is defined as follows,}
\begin{equation}
S_{ij}=-\frac{1}{8\pi }\left( \eta _{ij}-\delta _{ij}\eta _{kk}\right) .
\end{equation}
{Then, the discontinuity in the second fundamental form is expressed
as follows, }
\begin{equation}
\eta _{ij}=K_{ij}^{+}-K_{ij}^{-},
\end{equation}%
{where $K_{ij}$ describes the extrinsic curvature. The signs "$-$"
and "$+$" are related to the gravastar's interior and exterior regions,
respectively. The components of the extrinsic curvature tensor on both
surfaces of the shell region read}
\begin{equation}
K_{ij}^{\pm }=\left[ -n_{\nu }^{\pm }\left( \frac{\partial ^{2}x_{\nu }}{%
\partial \xi ^{i}\partial \xi ^{j}}+\Gamma _{\alpha \beta }^{\nu }\frac{%
\partial x^{\alpha }}{\partial \xi ^{i}}\frac{\partial x^{\beta }}{\partial
\xi ^{j}}\right) \right] ,
\end{equation}%
{where $\xi ^{i}$ represent the intrinsic coordinates on the shell,
and $-n_{\nu }^{\pm }$ are the normal unit $(n^{\nu }n_{\nu }=1)$ vectors}
\begin{equation}
n_{\nu }^{\pm }=\pm \left( {g^{\alpha \beta}\frac{\partial f}{\partial
x^{\alpha }}\frac{\partial f}{\partial x^{\beta }}}\right) ^{-\frac{1}{2}}%
\frac{\partial f}{\partial x^{\nu }}.
\end{equation}
{Now, from Lanczos equation spacetime, the stress-energy surface
tensor can be written as $S_{ij}=diag(\varrho ,P)$, where $\varrho$ is the
surface energy density and $P$ is the surface pressure, respectively.} So
according to the general formalism for three-dimensional spacetime \cite{51}
and using the above equations, and also by setting $r=D$, for uncharged
gravastars, we obtain
\begin{eqnarray}
\varrho _{uncharged}(D) &=&\frac{-1}{8\pi D}\left( \sqrt{-m_{0}-\frac{%
\Lambda \left( \varepsilon \right) D^{2}}{H^{2}(\varepsilon )}}-\sqrt{A+%
\frac{(\Lambda (\varepsilon )+4H_{0}^{2})D^{2}}{H^{2}(\varepsilon )}}\right)
, \\
&&  \notag \\
P_{uncharged}(D) &=&\frac{1}{8\pi }\left( \frac{-\frac{\Lambda \left(
\varepsilon \right) D}{H^{2}(\varepsilon )}}{\sqrt{-m_{0}-\frac{\Lambda
\left( \varepsilon \right) D^{2}}{H^{2}(\varepsilon )}}}-\frac{\frac{%
(\Lambda (\varepsilon )+4H_{0}^{2})D}{H^{2}(\varepsilon )}}{\sqrt{A+\frac{%
(\Lambda (\varepsilon )+4H_{0}^{2})D^{2}}{H^{2}(\varepsilon )}}}\right) .
\end{eqnarray}%
For charged gravastars, we have
\begin{eqnarray}
\varrho _{charged}(D) &=&\frac{-1}{8\pi D}\left( \sqrt{\mathcal{N}_{1}}-%
\sqrt{C_{0}+\mathcal{N}_{2}+\frac{\Lambda \left( \varepsilon \right) D^{2}}{%
H^{2}(\varepsilon )}}\right) , \\
&&  \notag \\
P_{charged}(D) &=&\frac{1}{8\pi }\left( \frac{-\frac{\Lambda \left(
\varepsilon \right) D}{H^{2}(\varepsilon )}-\frac{\frac{q^{2}L^{2}(%
\varepsilon )}{l^{2}}}{D}}{\sqrt{\mathcal{N}_{1}}}-\frac{\frac{\frac{16\pi
^{2}\sigma _{0}^{2}(m^{2}+2m+3)}{{(m+2)}^{2}(m+1)}D^{2m+3}+\Lambda \left(
\varepsilon \right) D}{H^{2}(\varepsilon )}}{\sqrt{C_{0}+\mathcal{N}_{2}+%
\frac{\Lambda \left( \varepsilon \right) D^{2}}{H^{2}(\varepsilon )}}}%
\right) ,
\end{eqnarray}%
where $\mathcal{N}_{1}$ and $\mathcal{N}_{2}$ are
\begin{eqnarray}
\mathcal{N}_{1} &=&-m_{0}+\frac{\Lambda \left( \varepsilon \right) D^{2}}{%
H^{2}(\varepsilon )}-\frac{2q^{2}L^{2}(\varepsilon )}{l^{2}}\ln \left( \frac{%
D}{l}\right) , \\
&&  \notag \\
\mathcal{N}_{2} &=&\frac{16\pi ^{2}\sigma _{0}^{2}(m^{2}+2m+3)D^{2m+4}}{{%
H^{2}(\varepsilon )(m+2)}^{3}(m+1)}.
\end{eqnarray}%
It is notable that $\varrho _{uncharged}$ and $P_{uncharged}$ are related to
the line energy density and line pressure of uncharged gravastar in
gravity's rainbow, respectively. Also, $\varrho _{charged}$ and $P_{charged}$%
, are related to the line energy density and the line pressure of charged
gravastar in this gravity, respectively. We have drawn our results for the
line energy density and line pressure versus radius $D$ for both state of
charged and uncharged in Figs. \ref{rhounch}-\ref{Pch}. It is clear that the
line energy density for both state of charged and uncharged gravastars
increases (see Figs. \ref{rhounch} and \ref{rhoch}). But the line pressure
for both state of charged and uncharged gravastars decreases as radius $D$
increases (see Figs. \ref{Punch} and \ref{Pch}).

\subsection{Equation of State of thin-Shell gravastars in gravity's rainbow}

\label{section.Equation of State} 
\begin{figure}[tbp]
\centering
\includegraphics[scale=0.3]{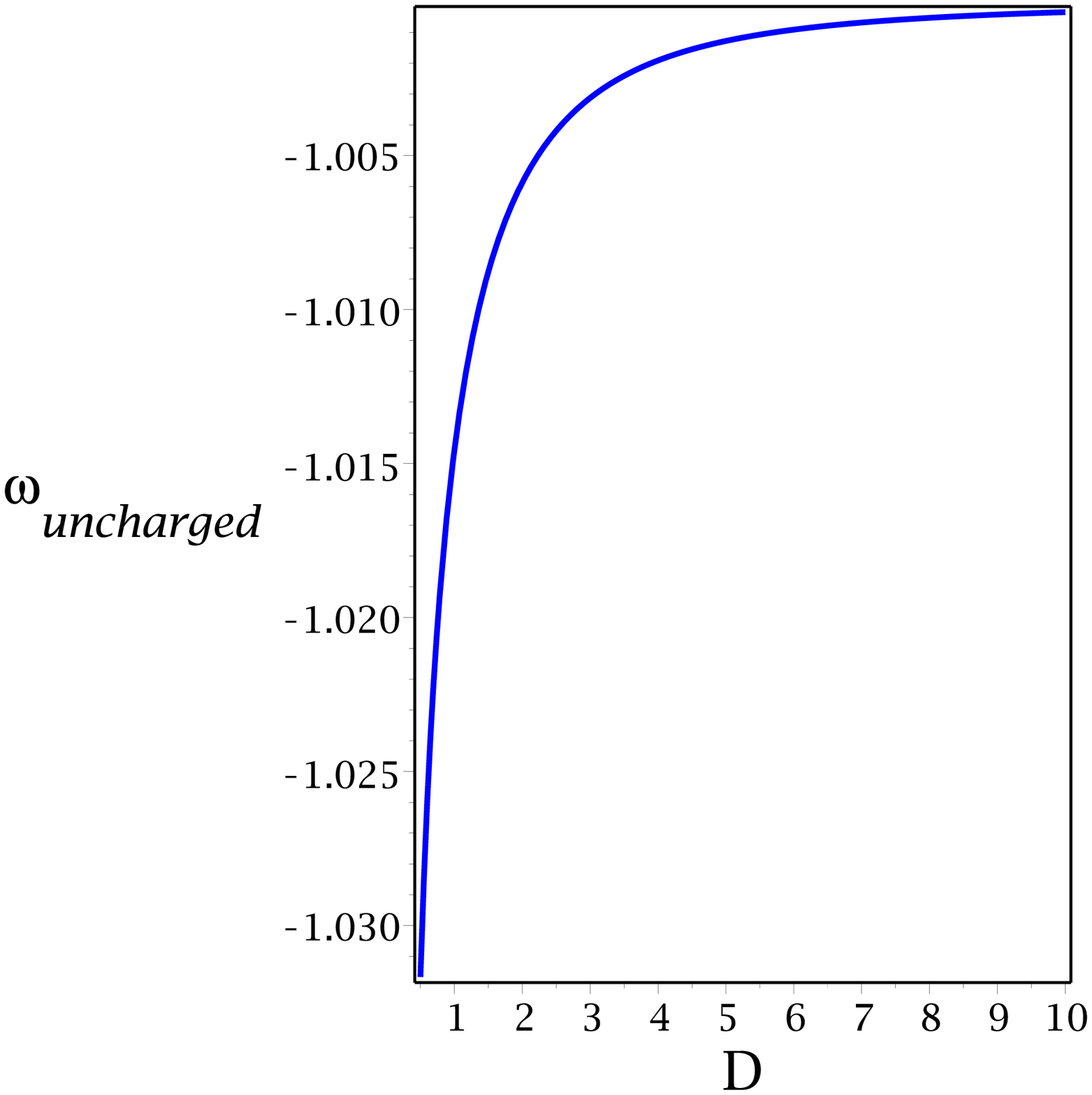}
\caption{$\protect\omega _{uncharged}$ vs radius $D$. We have chosen $%
m_{0}=-4$, $H_{0}=70$, $A=1$, $\Lambda (\protect\varepsilon )=-1$, and $H(%
\protect\varepsilon )=1.1$. }
\label{omegauncharged}
\end{figure}

{The equation of state parameter $%
\omega (r)$ (at $r=D$) is described as follows}
\begin{equation}
\omega (D)=\frac{P(D)}{\varrho (D).}
\end{equation}
{Therefore, using equations for line energy density and line
pressure, we can obtain the equation of state parameter in the following
form,}
\begin{equation}
\omega _{uncharged}(D)=\frac{\frac{\frac{\Lambda \left( \varepsilon \right)
D^{2}}{H^{2}(\varepsilon )}}{\sqrt{-m_{0}-\frac{\Lambda \left( \varepsilon
\right) D^{2}}{H^{2}(\varepsilon )}}}+\frac{\frac{(\Lambda (\varepsilon
)+4H_{0}^{2})D^{2}}{H^{2}(\varepsilon )}}{\sqrt{A+\frac{(\Lambda
(\varepsilon )+4H_{0}^{2})D^{2}}{H^{2}(\varepsilon )}}}}{\sqrt{-m_{0}-\frac{%
\Lambda \left( \varepsilon \right) D^{2}}{H^{2}(\varepsilon )}}-\sqrt{A+%
\frac{(\Lambda (\varepsilon )+4H_{0}^{2})D^{2}}{H^{2}(\varepsilon )}}},
\end{equation}%
{for charged}
\begin{equation}
\omega _{charged}(D)=\frac{\frac{-\frac{\Lambda \left( \varepsilon \right)
D^{2}}{H^{2}(\varepsilon )}-\frac{q^{2}L^{2}(\varepsilon )}{l^{2}D}}{\sqrt{%
\mathcal{N}_{1}}}-\frac{\frac{\frac{16\pi ^{2}\sigma _{0}^{2}(m^{2}+2m+3)}{{%
(m+2)}^{2}(m+1)}D^{2m+4}+\Lambda \left( \varepsilon \right) D^{2}}{%
H^{2}(\varepsilon )}}{\sqrt{C_{0}+\mathcal{N}_{2}+\frac{\Lambda \left(
\varepsilon \right) D^{2}}{H^{2}(\varepsilon )}}}}{\sqrt{\mathcal{N}_{1}}-%
\sqrt{C_{0}+\mathcal{N}_{2}+\frac{\Lambda \left( \varepsilon \right) D^{2}}{%
H^{2}(\varepsilon )}}},
\end{equation}%
{in which $\omega _{uncharged}$, and $\omega _{charged}$ are the
equation of state parameters for uncharged and charged gravastars. We have
plotted the equation of state parameters $\omega_{uncharged}$, and $\omega
_{charged}$ vs radius $\mathbf{D}$ in Fig. \ref{omegauncharged} and Fig. \ref%
{omegacharged}, respectively.}
\begin{figure}[tbp]
\centering
\includegraphics[scale=0.3]{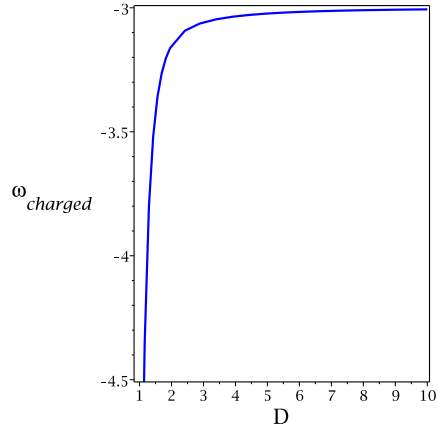}
\caption{$\protect\omega _{charged}$ vs radius $D$. We have chosen $m_{0}=-4$%
, $L(\protect\varepsilon )=1.1$, $H(\protect\varepsilon )=1.1$, $\protect%
\sigma _{0}=4$, $C_{0}=1$, $\Lambda (\protect\varepsilon )=-1$, $m=1$, $q=1$%
, and $l=1$.}
\label{omegacharged}
\end{figure}
{From these figures, we see that the equation of state parameter
increase as the radius increases, but it keeps a negative sign for both
charged and uncharged states.}

\subsection{Stability of thin-Shell gravastars in gravity's rainbow}

\label{V} \label{section.stability}

In this section, we want to study the stability of thin-shell gravastars for
three-dimensional uncharged and charged AdS in gravity's rainbow. The stable
and unstable formations of thin-shell gravastars can be analyzed through the
behavior of $\eta $ as an effective parameter in determining the stability
regions of the respective solutions that defined as follows \cite{48},
\begin{equation*}
\eta (r)=\frac{P^{\prime }(r)}{\varrho ^{\prime }(r)}.
\end{equation*}
The parameter $\eta $ can be interpreted as the squared speed of sound
satisfying $0\leq \eta \leq 1$ commonly. It is obvious that the speed of
sound should not exceed the speed of light. But, to check the stability of
the gravastar, this limitation may not be satisfied on the surface layer
\cite{49,50}. The parameter $\eta $ for uncharged gravastar in $r=D$ can be
written in the following form,
\begin{equation}
\eta _{uncharged}(D)=\frac{-\frac{\Lambda \left( \varepsilon \right)
^{2}D^{2}}{\mathcal{N}_{3}^{\frac{3}{2}}H^{4}(\varepsilon )}-\frac{\Lambda
\left( \varepsilon \right) }{\sqrt{\mathcal{N}_{3}}H^{2}(\varepsilon )}+%
\frac{(\Lambda \left( \varepsilon \right) +4H_{0}^{2})^{2}D^{2}}{\mathcal{N}%
_{4}^{\frac{3}{2}}H^{4}(\varepsilon )}-\frac{\Lambda \left( \varepsilon
\right) +4H_{0}^{2}}{\sqrt{\mathcal{N}_{4}}H^{2}(\varepsilon )}}{\frac{\sqrt{%
\mathcal{N}_{3}}-\sqrt{\mathcal{N}_{2}}}{D^{2}}+\frac{\frac{\Lambda \left(
\varepsilon \right) D}{\sqrt{\mathcal{N}_{3}}H^{2}(\varepsilon )}+\frac{%
(\Lambda \left( \varepsilon \right) +4H_{0}^{2})D}{\sqrt{\mathcal{N}_{4}}%
H^{2}(\varepsilon )}}{D}},
\end{equation}%
where $\mathcal{N}_{3}$ and $\mathcal{N}_{4}$ are
\begin{eqnarray}
\mathcal{N}_{3} &=&-m_{0}-\frac{\Lambda \left( \varepsilon \right) D^{2}}{%
H^{2}(\varepsilon )}, \\
&&  \notag \\
\mathcal{N}_{4} &=&A+\frac{(\Lambda (\varepsilon )+4H_{0}^{2})D^{2}}{%
H^{2}(\varepsilon )}.
\end{eqnarray}
Therefore, we analyze the stability of gravastar configurations by analyzing
the sign of the parameter $\eta $. Because the expression is complicated, we
use the graphical behavior of $\eta _{uncharged}(D)$ in Fig. \ref{stability
uncharged}. In this figure, we examine that thin-shell of three-dimensional
uncharged gravastars in gravity's rainbow expresses stable behavior for very
values of $D$.
\begin{figure}[tbp]
\centering
\includegraphics[scale=0.3]{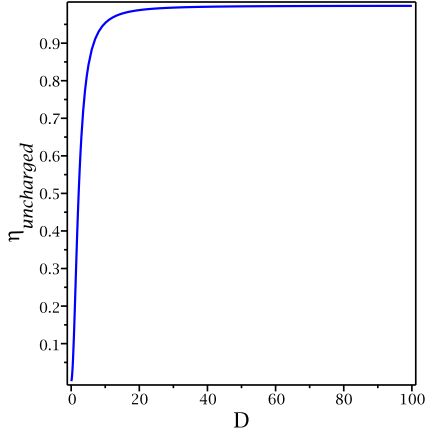}
\caption{$\protect\eta _{uncharged}$ vs radius $D$. We have chosen $%
m_{0}=-80 $, $H_{0}=70$, $A=1$, $\Lambda (\protect\varepsilon )=-1$, and $H(%
\protect\varepsilon )=1.1$. }
\label{stability uncharged}
\end{figure}
We obtain the parameter $\eta $ for charged garavastars in the following
form
\begin{equation}
\eta _{charged}(D)=\frac{\left( \frac{\frac{-\Lambda \left( \varepsilon
\right) }{H^{2}(\varepsilon )}+\frac{L^{2}(\varepsilon )q^{2}}{l^{2}D^{2}}}{%
\sqrt{\mathcal{N}_{1}}}-\frac{\left( \frac{\Lambda \left( \varepsilon
\right) D}{H^{2}(\varepsilon )}+\frac{q^{2}L^{2}(\varepsilon )}{l^{2}D}%
\right) \left( \frac{2\Lambda \left( \varepsilon \right) D }{%
H^{2}(\varepsilon )}+\frac{2q^{2}L^{2}(\varepsilon )}{l^{2}D}\right) }{2%
\mathcal{N}_{1}^{\frac{3}{2}}}-\frac{\frac{(2m+3)(m+2)\mathcal{N}_{2}}{D^{2}}%
+\Lambda \left( \varepsilon \right) }{\sqrt{C_{0}+\mathcal{N}_{2}+\Lambda
\left( \varepsilon \right) D^{2}}}+\mathcal{N}_{5}\right) }{\frac{\sqrt{%
\mathcal{N}_{1}}-\sqrt{C_{0}+\mathcal{N}_{2}+\Lambda \left( \varepsilon
\right) D^{2}}}{D^{2}}+\frac{\frac{\frac{\Lambda \left( \varepsilon \right) D%
}{H^{2}(\varepsilon )}+\frac{L^{2}(\varepsilon )q^{2}}{l^{2}D}}{\sqrt{%
\mathcal{N}_{1}}}+\frac{\frac{2\mathcal{N}_{2}}{D}+\frac{m\mathcal{N}_{2}}{D}%
+\Lambda \left( \varepsilon \right) D}{\sqrt{C_{0}+\mathcal{N}_{2}+\Lambda
\left( \varepsilon \right) D^{2}}}}{D}},
\end{equation}%
where $\mathcal{N}_{5}$\ is the following function,
\begin{equation}
\mathcal{N}_{5}=\frac{\left( \frac{(m+2)\mathcal{N}_{2}}{D}+\Lambda \left(
\varepsilon \right) D\right) \left( \frac{2(m+2)\mathcal{N}_{2}}{D}+2\Lambda
\left( \varepsilon \right) D\right) }{2\left( {{C_{0}+\mathcal{N}%
_{2}+\Lambda \left( \varepsilon \right) D}^{2}}\right) ^{3/2}}.  \notag
\end{equation}
We study the graphical behavior of $\eta _{charged}(D)$ in Fig. \ref%
{stability charged}. This figure examines how the thin shell of
three-dimensional charged gravastars in gravity's rainbow expresses stable
behavior. As one can see, the charged three-dimensional gravastars in
gravity's rainbow are stable object.
\begin{figure}[tbp]
\centering
\includegraphics[scale=0.3]{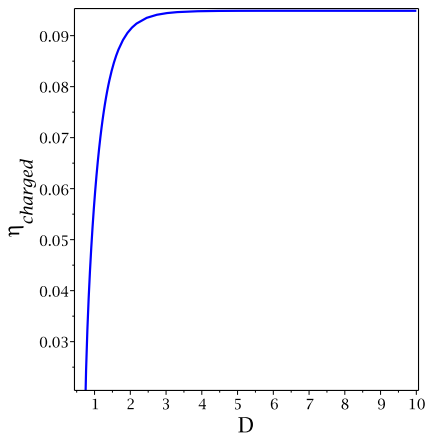}
\caption{$\protect\eta _{charged}$ vs radius $D$. We have chosen $m_{0}=-80$%
, $L(\protect\varepsilon )=1.1$, $H(\protect\varepsilon )=1.1$, $\protect%
\sigma _{0}=4$, $C_{0}=1$, $\Lambda (\protect\varepsilon )=-1$, $m=1$, $q=1$%
, and $l=1$.}
\label{stability charged}
\end{figure}

\section{CONCLUSION}

\label{VI}

\label{section.conc}

%In this paper, we studied the gravastars in the contexts of both three-dimensional AdS spacetime and gravity's rainbow theory.
In fact, the gravastar is the short form of Gravitationally vacuum stars
which has given a new idea in the gravitational system. Such a stellar model
{which is free from any singularity}, can be considered an alternative to
black holes. Gravastar can be described through the interior region, thin
shell region, and exterior region. In this paper, we studied the gravastars
in the contexts of both three-dimensional AdS spacetime and gravity's
rainbow theory. We discussed the geometry of the uncharged gravastar model
in gravity's rainbow. In the interior region, the gravastar follows the
equation of state (EoS) $p=-\rho $. Some physical quantities such as energy
density, pressure and gravitational mass were obtained for the interior
region. In the exterior region, we found the vacuum exterior region whose
EoS is given by $p=\rho =0$. The solution corresponds to a static BTZ black
hole in the gravity's rainbow. In the shell region, the fluid source follows
the EoS $p=\rho $. In this region, we considered the thickness the shell of
the gravastar to be very small, because we assumed that the interior and
exterior regions join together at a location. Employing the approximation $%
0<g(r)\equiv h<<1$, we found the analytical solutions within the thin shell
of the uncharged gravastar. Next, we calculated the physical parameters such
as the proper length of the thin shell, energy, and entropy inside the thin
shell of the uncharged gravastar, and showed that these parameters are
directly proportional to the proper thickness of the shell ($\delta $) due
to the approximation $\delta <<1$. Also, we investigated the physical
parameters such as the proper length $l$, the energy $E$, and the entropy $S$
versus the rainbow function $H(\varepsilon )$. Here, we saw that all three
parameters of the shell of the uncharged gravastar decreasing function of
rainbow function $H(\varepsilon )$. In the next section, While considering a
new model of charged gravastar in connection to the electrovacuum exterior
three-dimensional AdS spacetime in the context of gravity's rainbow theory,
we investigated the role of electromagnetic field on an isotropic stellar
model. To do this, we first wrote the Einstein-Maxwell's field equations in
the framework of the gravity's rainbow. Then, we obtained the geometry of
charged gravastar model. Some physical quantities such as energy density,
pressure and gravitational mass were obtained for the interior region. In
the exterior region, we found the electrovacuum exterior region. Solution
corresponds to a static, charged BTZ black hole in the gravity's rainbow. In
the shell region, under approximation $0<g(r)\equiv h<<1$, we obtained the
analytical solutions within the thin shell of the charged gravastar. We
computed the physical parameters such as the proper length of the thin
shell, energy, and entropy inside the thin shell of the charged gravastar,
and showed that these parameters are directly proportional to the proper
thickness of the shell ($\delta $) due to the approximation $\delta <<1$.
These physical parameters significantly depend on the rainbow function $%
H(\varepsilon )$. From the results of calculations, we saw that the proper
length of the shell, energy, and entropy inside the shell of the charged
gravastar increase as thickness increases. Also, the proper length decreases
as the radius increases. The energy and entropy inside the shell always
increase as the radius increases. Besides, by using the Darmois-Israel
formalism, the matching between the surfaces of interior and exterior
regions of the gravastars were studied. According to the matching
conditions, the line energy density and the line pressure were obtained for
both charged and uncharged states. Our results show that the line density
increase as the radius increases, but the line pressure always decreases as
the radius increases, but keeps the negative sign for both charged and
uncharged states. {Also, the equation of state parameter on the
surface was obtained which shows that the equation of state parameter
increase by increasing the radius. Here we saw that there is a negative sign
for both charged and uncharged states.} Finally, we explored the stable
regions of uncharged and charged gravastar in gravity's rainbow.
%%%%%%%%%%%%%%%%%%%%%%%%%%%%%%%%%%%%%%%%%%%%%%%%%%%%%%%%%%%%%%%%%%%%%%%%%%%%%%%%%%%%%%%%

\acknowledgements{G. H. Bordbar wishes to thank the Shiraz University Research Council.
B. Eslam Panah thanks the University of Mazandaran. The
University of Mazandaran has supported the work of B. Eslam Panah by title
"Evolution of the masses of celestial compact objects in various gravity".}

%%%%%%%%%%%%%%%%%%%%%%%%%%%%%%%%%%%%%%%%%%%%%%%%%%%%%%%%%%%%%%%%%%%%%%%%%%%%%%%%%%%%

\end{document}